\documentclass[aps, prl, reprint, groupedaddress, superscriptaddress, footinbib, floatfix]{revtex4-1}

\usepackage{amsmath, amssymb, bm}
\usepackage{graphicx}
\usepackage{natbib}
\usepackage{soul}
\usepackage{float}
\usepackage{upgreek}
\usepackage[euler]{textgreek}
\usepackage[english]{babel} 
\usepackage{blindtext}
\usepackage[dvipsnames]{xcolor}
\usepackage{braket}
\usepackage{hyperref}
\definecolor{ddblue}{RGB}{0,0,160}
\hypersetup{colorlinks=true,linktoc=all,linkcolor=ddblue,citecolor=blue,urlcolor=[rgb]{0, 0.38, 0}}

\newcommand*\vecc[1]{\ifmmode\bm{#1}\else\textbf{#1}\fi}

\begin{document}
	
	\preprint{APS/123-QED}
	
	\title{Sub-Natural Linewidth Superradiant Lasing with Cold $^{88}$Sr Atoms}
	
	\author{Sofus Laguna Kristensen}
	\email{sofus.kristensen@nbi.ku.dk}
	\author{Eliot Bohr}
	\author{Julian Robinson-Tait}
	\affiliation{Niels Bohr Institute, University of Copenhagen, DK-2100 Copenhagen, Denmark}
	
	\author{Tanya Zelevinsky}
	\affiliation{Department of Physics, Columbia University, 538 West 120th Street, New York, NY 10027-5255, USA}
	
	\author{Jan W. Thomsen}
	\author{Jörg Helge Müller}
	\email{muller@nbi.ku.dk}
	
	\affiliation{Niels Bohr Institute, University of Copenhagen, DK-2100 Copenhagen, Denmark}
	
	\date{\today}
	
	\begin{abstract}
		Superradiant lasers operate in the bad-cavity regime, where the phase coherence is stored in the spin state of an atomic medium rather than in the intracavity electric field. Such lasers use collective effects to sustain lasing and could potentially reach considerably lower linewidths than a conventional laser. Here, we investigate the properties of superradiant lasing in an ensemble of ultracold $^{88}$Sr atoms inside an optical cavity. We extend the superradiant emission on the $7.5$ kHz wide $^3P_1$  $\rightarrow$ $^1S_0$ intercombination line to several milliseconds, and observe steady parameters suitable for emulating the performance of a continuous superradiant laser by fine tuning the repumping rates. We reach a lasing linewidth of 820 Hz for 1.1 ms of lasing, nearly an order of magnitude lower than the natural linewidth. 
	\end{abstract}

	\maketitle
Lasers emitting narrow frequency spectra have led to numerous advances in precision-measurement technologies and fundamental research. Optical atomic clocks, which rely on narrow linewidth and ultrastable lasers, have achieved fractional instabilities down to a few parts in $10^{19}$ \cite{campbell2017,brewer2019, McGrew2018,Bloom2014,Ushijima2015,Peik2016}. However, such lasers require a separate high-finesse cavity and the stability of atomic clocks is currently limited by the duty cycle of the atomic interrogation. Approaches to bring down the dead time between the measurements include using multiple atomic ensembles to bridge the time gap \cite{Schioppo2017,soerensen2013,Kim2023,li2022}, non-destructive measurements of the atomic ensemble \cite{Hobson2019,Vallet2017}, or using a conveyor-belt scheme for continuous loading and interrogation of atoms \cite{Katori2021}. Another proposal to bring down the dead time is to use the coupling between a cavity and a cold atom sample to conduct passive cavity-assisted nonlinear spectroscopy \cite{Martin2011,Westergaard2015,Christensen2015} for laser stabilization, but it requires fine tuning of the feedback bandwidth and meticulous management of technical noise. 

Recently, interest has grown in developing an active atomic clock \cite{schaffer2020,Bohnet2012ASS,norcia2016,norcia2018frequency, Haake1993} that removes the need for a separate high-finesse cavity for short-term stabilization of a reference laser. In an active atomic clock, atoms emitting coherently on a narrow transition into the mode of an optical cavity constitute the frequency reference \cite{mHzlaser,Chen2009}. By operating in the bad-cavity regime, where the lasing frequency is dictated by the linewidth of the atomic ensemble rather than the cavity, it is possible to significantly reduce frequency fluctuations of the laser due to thermal noise in the cavity mirrors, which is a limitation in current state-of-the-art clocks \cite{Numata2004,Cole2013,Ludlow2015}. 

Several studies investigating pulsed superradiant light \cite{norcia2018frequency,tang2021,schaffer2020} depend on single excitation pulses to generate superradiant bursts. However, the collectively enhanced emission rate inherently produces pulses with a shorter duration than the natural lifetime of the respective transition, resulting in Fourier limited spectral features of the emitted light. Seminal work in \cite{norcia2016} demonstrated extended superradiant pulses by repumping of the atoms and achieved 100s of $\mu$s of lasing on the kHz transition in $^{\text{88}}$Sr. 

In this Letter, we present a lasing scheme on the $689$ nm $^3P_1$, $m_j=0$  $\rightarrow$ $^1S_0$ intercombination transition which has a natural linewidth $\gamma_l=2\pi \times 7.5$ kHz. Adjustable repumping rates allow us stabilize lasing parameters and approach the frequency behavior of a truly continuous superradiant laser. We use a free-falling cloud of ultracold $^{\text{88}}$Sr atoms coupled to a low finesse cavity to produce superradiant lasing pulses. We achieve stable lasing power output and extend the emission duration by shelving atoms in the metastable $^3P_2$ level to distribute the available gain over a longer time. With this scheme, we maximize the length of the pulses in order to analyze their frequency content and achieve more than 1 ms of steady-state lasing with several $10^{-9}$ watts average output power. We push the Fourier limit almost an order of magnitude below the natural linewidth of the transition, and demonstrate the feasibility of active lasing schemes for future superradiant atomic clocks.

\begin{figure}[t!]
\centering
\includegraphics[width=1\linewidth]{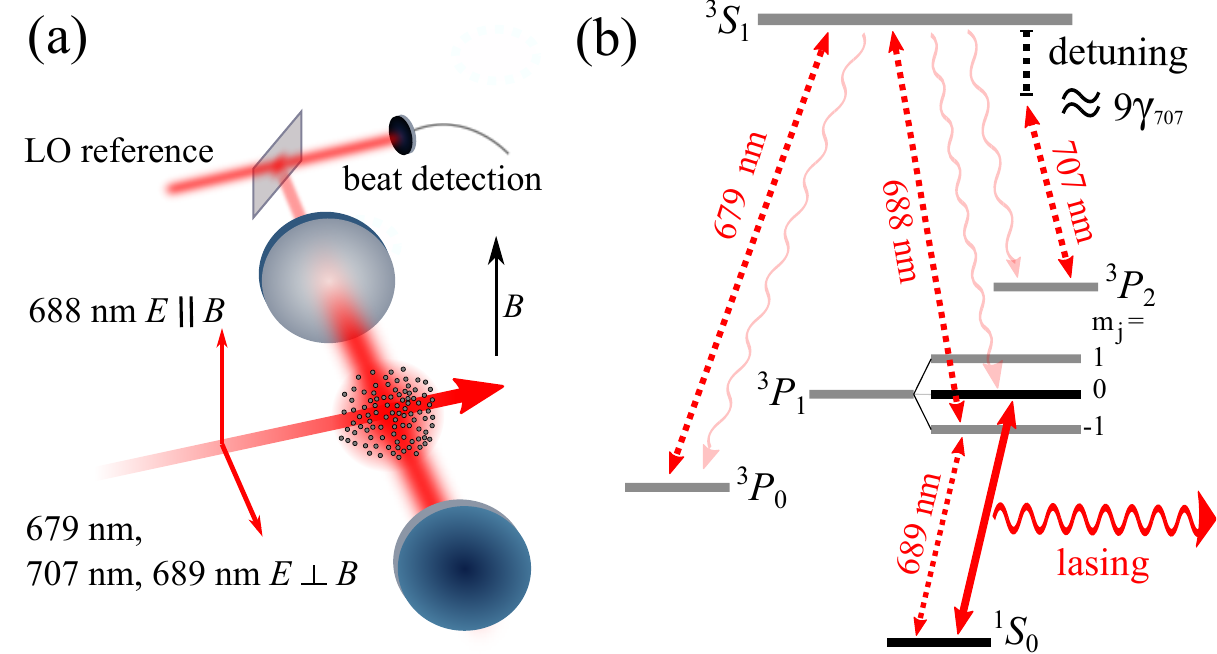} 
\caption{(a) Experimental diagram. Repump laser beams at 679 nm, 707 nm, 688 nm, and 689 nm are incident perpendicular to the cavity. The polarizations of the 688 nm and 689 nm repump light are chosen to avoid exciting or depleting $^3P_1$, $m_j=0$, and to limit light shifts of the lasing transition. We beat the emitted light from the cavity against a stable reference laser to measure the frequency stability of the emitted pulse. (b) Level scheme of $^{\text{88}}$Sr showing the lasing (solid arrow), the repumping (dotted arrows) and the spontaneous decay (wavy arrows). We detune the 707 nm repump light by $2\pi \times 60 \text{ MHz}\approx 9\gamma_{707}$ to shelve atoms in $^3P_2$ during lasing.} 
\label{fig1}
\end{figure}

Our experimental system consists of $N=4\times10^7$ $^{\text{88}}$Sr atoms cooled to 2 $\mu$K using 461 nm and 689 nm magneto-optical trap (MOT) stages. The MOT is positioned using bias magnetic fields to overlap with the fundamental mode of a horizontal 21 cm long optical cavity, as depicted in Fig. \ref{fig1}(a). The cavity has a finesse of $10^3$, a linewidth $\kappa = 2\pi \times 780$~kHz, and free spectral range (FSR) of $2\pi \times 781$ MHz. The radii of curvature of the mirrors are 9 m and the cavity mode waist radius is $450$~$\mu$m. The cavity length is locked with light one FSR away from atomic resonance to ensure that the locking light does not significantly interact with the atoms. The diameter of the $689$~nm MOT in the end of the cooling cycle is 100 $\mu$m, such that the atomic ensemble fits well within the TEM00 cavity mode volume. 
The atoms are distributed across the nodes and antinodes and therefore have inhomogeneous couplings to the cavity mode, $g_0$.  We operate in the single-atom weak coupling regime, with Purcell factor \cite{TANJISUZUKI2011_2} $C = \frac{4g_0^2}{\kappa\gamma_l}= 4.4 \times 10^{-4}$, but in the collective strong coupling regime, $NC \gg 1$.  This is verified by a resolved normal-mode splitting measurement of $\Omega_N = 2\pi \times 5.7$ MHz, which shows that the collective vacuum Rabi frequency exceeds the other relevant decoherence rates in the system: $\Omega_N \gg \kappa,\gamma_l$. After cooling the atoms, we switch off the MOT laser beams and magnetic coils, leaving only a vertical bias field of $2$~Gauss. This bias field provides a quantization axis and leads to a splitting of 4.2 MHz between each of the $^3P_1$ magnetic sublevels. 

By turning on 4 repump lasers (dotted arrows in Fig. \ref{fig1}(b)) we incoherently populate $^3P_1$, $m_j=0$ which results in superradiant laser emission. The polarization of the 689 nm repump light is oriented perpendicularly to the bias field, allowing only transitions to $^3P_1$, $m=\pm1$ states with a detuning chosen to address only one state. The 688 nm repump light polarization is parallel to the bias field to drive $\Delta m_j =0$ transitions between $^3P_1$, $m_j=\pm1$ and $^3S_1$, $m_j=\pm1$, while not depleting the upper lasing level, $^3P_1$, $m_j=0$ state, due to selection rules. Two more repump lasers at 679 nm and 707 nm counteract the branching decay from $^3S_1$ into the $^3P_0$ and $^3P_2$ levels. The 688 nm, 679 nm and 707 nm repump lasers all operate close to saturation intensity of the respective transitions. It is important that the atoms reach the upper lasing level via spontaneous decay to ensure that we do not have any coherence between the repump lasers and the emitted light. Any coherence could introduce unwanted phase noise.

All repump lasers are copropagating and oriented perpendicularly to the cavity axis, as shown in Fig. \ref{fig1}(a). This ensures that the Doppler shift induced by repumping is smallest for light emitted along the cavity axis, and it maximizes the time where the Doppler width $\gamma_{D}$ of the atomic ensemble (which is the dominant broadening) is much narrower than the cavity linewidth, such that we operate in the bad-cavity regime $\kappa / \gamma_{D} \gg 1$ throughout the lasing emission. The copropagation of the repump lasers also limits the variable light shifts (primarily from the 689 nm and 688 nm lasers) since the atoms stay well within the Gaussian intensity profiles of the repump beams, even as the cloud is accelerated by radiation pressure. 

We estimate from numerical simulations that for each emitted photon in the cavity mode an atom on average absorbs 9 repump photons and spontaneously emits 8. The Doppler shift from the absorbed repump photons for one lasing excitation cycle is $\sim$85 kHz, more than an order of magnitude larger than $\gamma_l$. As the atoms are repumped and accelerate out of the cavity mode, they experience a 689 nm repump laser frequency with a detuning that changes for each repumping cycle. When all repump lasers are on resonance, the high repumping rate into the upper lasing level results in a collectively enhanced emission rate that exceeds the repumping rate of the 689 nm repump laser $\Omega_{N} > \Gamma_{689}$ due to our high intracavity atom number.

This unstable balance between the repumping rate and the collectively enhanced emission rate causes the amplitude of the collective dipole to oscillate in time, producing a spiked behavior in the lasing intensity. The high repumping rate quickly heats and pushes the cloud out of the cavity mode which terminates the lasing after only a few hundred $\mu$s. By carefully adjusting the detuning of the 707 nm repump light, we can transition from an unstable lasing regime with oscillatory output power towards a steady-state lasing regime. 

In the steady-state regime we reduce our repumping rate out of $^3P_2$ by detuning the 707 nm repump laser by $2\pi \times 60$ MHz (about 9 times the linewidth of the 707 nm transition, $\gamma_{707}$) from the $^3P_2$, $m_j=0$ $\rightarrow$ $^3S_1$, $m_j=0$ transition. This shelves a significant fraction of the atoms in the $^3P_2$ level during lasing, which in turn reduces the collective coupling of the atom-cavity system, leading to a lower but more stable output power of the emission. Additionally, the heating and deflection of the atom cloud induced by repumping is reduced, allowing for longer lasing pulses. The repumping scheme grants independent tuning of the ground state depumping. We can lower the power of the 689 nm repump laser down to a similiar magnitude as the saturation intensity, which limits power broadening of the lasing transition as the repumping from the ground state only needs to be strong enough to sustain population inversion on the lasing transition, $\Gamma_{689} \ge \Omega_N$. 

\begin{figure}[!t]
\centering
\includegraphics[width=1\linewidth]{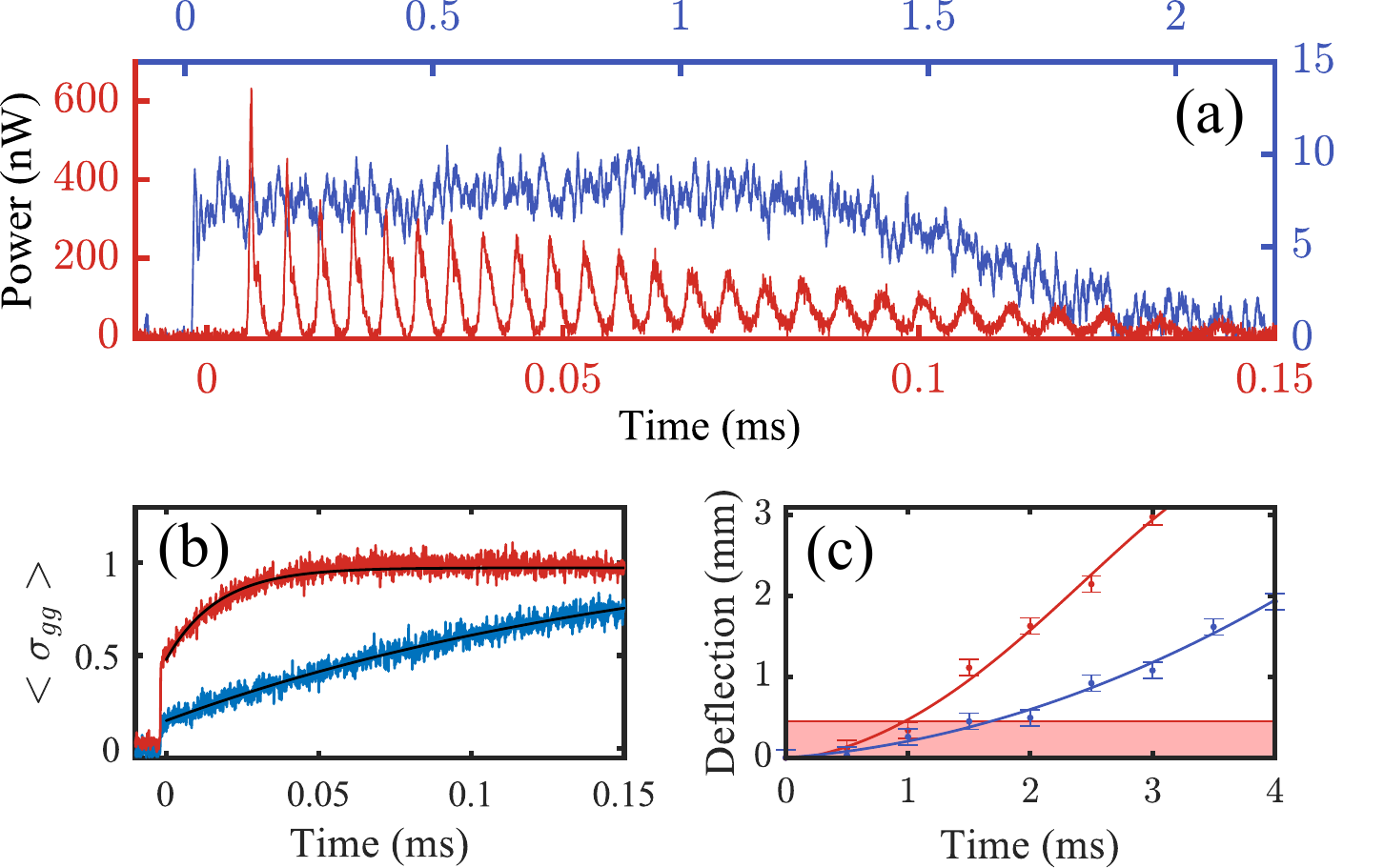}
\caption{Two different regimes of lasing, governed by the repumping rate. (a) In blue, on the right and top axis, a pulse with the 707 nm repump laser detuned by $2\pi \times 60$ MHz. With this detuning, the atoms are shelved in the $^3P_2$ state for 100s of $\mu$s leading to a pulse with steady lasing intensities for over a ms. In comparison, on the left and bottom axis, in red, we show a pulse with the 707 nm repump laser tuned to resonance. This pulse has a high peak output power of 600 nW, but only lasts for 100 $\mu$s.
(b) Fluorescence measurement of the ground state atoms after turning off the 689 nm repump laser immediately after 100 $\mu$s of repumping for the oscillatory (red) and steady-state (blue) regimes. Exponential fits (in black) yield the characteristic time constants of $20.4(4)$ $\mu$s (oscillatory regime) and $190(19)$ $\mu$s (steady-state regime). (c) The mechanical deflection of the MOT cloud measured by shadow imaging the atoms after a variable time-of-flight. The shaded red area indicates the waist radius of the cavity. The two different sets of data points correspond to the pulses shown in (a) in red and blue.}
\label{fig2}
\end{figure}

Figure \ref{fig2}(a) shows the two lasing regimes. On the left $y$-axis and bottom $x$-axis (in red) we show lasing where the 707 nm repump light is on resonance. In this regime, we can achieve peak powers exceeding 600 nW. On the right $y$-axis and top $x$-axis (in blue) the 707 nm repump laser is detuned by $2\pi \times 60$ MHz, which generates less intense but longer pulses with regions of steady-state lasing up to 1.5 ms. 

To quantify the effect of detuning the 707 nm repump laser on the population dynamics, we compare the repumping rates for our two lasing regimes in Fig. \ref{fig2}(b). We apply all of the repump lasers for 100 $\mu$s, after which we turn off the 689 nm repump laser which ends the superradiant emission as the population inversion is not sustained. Immediately after, we turn on the 461 nm MOT light and measure via fluorescence the proportion of atoms that returns to the ground state, $\langle \sigma_{gg} \rangle$, as a function of time. 

In the oscillatory lasing regime, we find a characteristic time for the atoms to return to the ground state $\tau = 20.4(4)$ $\mu$s which is consistent with the free-space natural decay from $^3P_1$. For the steady-state regime, the time constant $\tau = 190(19)$ $\mu$s is an order of magnitude larger. We observe a step in fluorescence at the onset of the measurement. This step provides an insight into the number of atoms in the ground and excited state. For the oscillatory regime this means that almost all of the atoms are either in $^3P_1$, $m_j=0$ or in the ground state. For the steady-state regime we observe only 10-15\% of the atoms in the ground state initially, i.e. most of the 4 $\times$ $10^{7}$ atoms in the cloud are shelved in $^3P_2$ during lasing, and are only slowly repumped back into the upper lasing level to sustain the lasing. 
\begin{figure}[!t]
\centering
\includegraphics[width=1\linewidth]{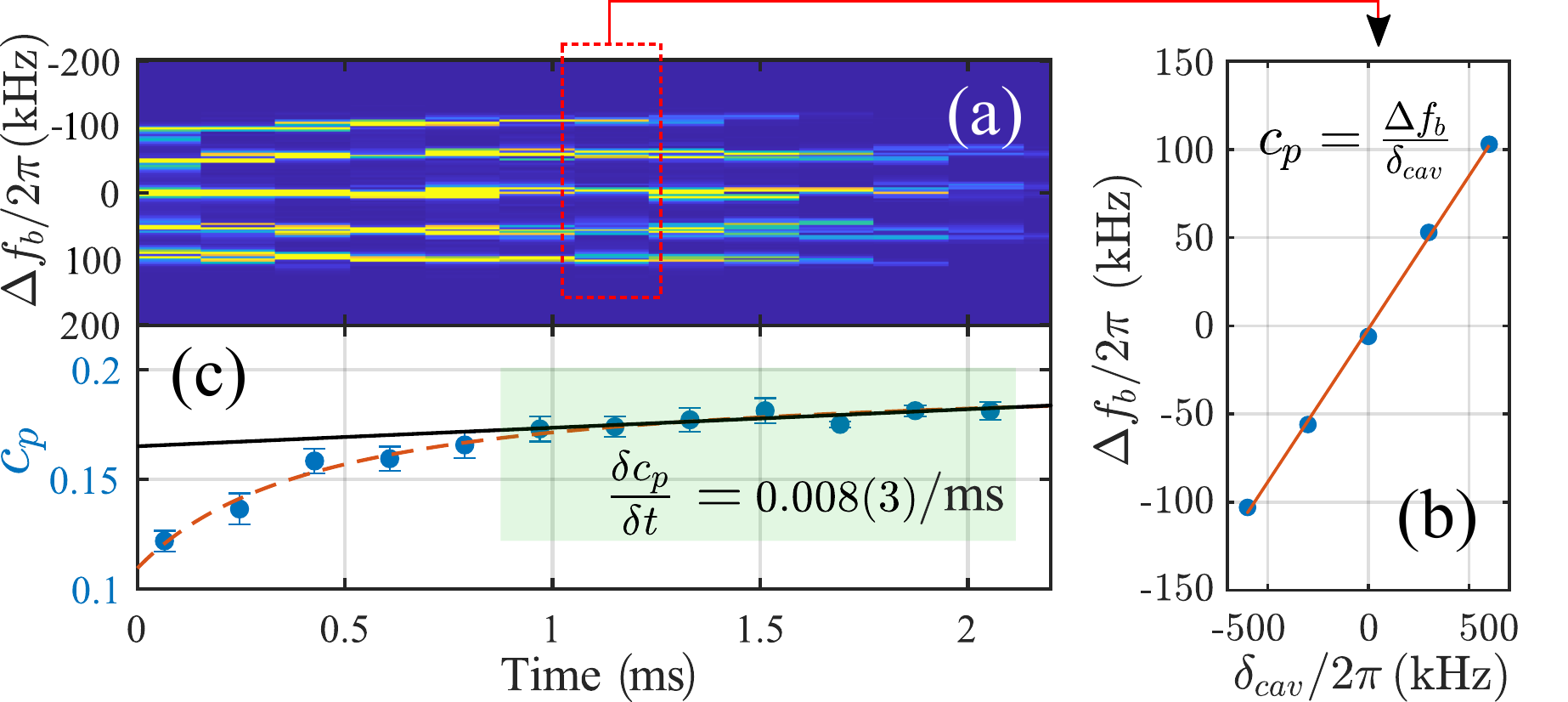} 
\caption{(a) Combined spectrogram of five pulses with different cavity detunings, $\delta_{cav}/2\pi$, from $-600$ kHz to $600$ kHz in steps of $300$ kHz. (b) We find the time-dependent $c_p$ from the slope of a straight line fit between the cavity offset and the lasing frequency. (c) The calculated cavity pulling $c_p(t)$. We observe a region where $\Delta c_p(t)=0.008(3)$ for $\Delta t=1$ ms (black solid line fit).}
\label{fig3}
\end{figure}
In Fig. \ref{fig2}(c) we measure the physical deflection of the atom cloud due to photon recoils in the two lasing regimes. Each set of data points, red and blue, correspond to the regimes presented in Fig. \ref{fig2}(a) and the repumping rates in Fig. \ref{fig2}(b). We fit high-order polynomials as guides for the eye. The shaded red area indicates the cavity waist radius. In the steady-state regime, the weakened repumping rate reduces the physical deflection of the atoms, allowing for a longer interaction with the cavity. Here, the lasing terminates when the cloud leaves the cavity mode volume and the collective coupling goes below its lasing threshold. For the oscillatory regime, the acceleration of the cloud quickly detunes the 689 nm repump transition such that the repumping can no longer sustain population inversion, terminating the lasing while the cloud is still within the cavity mode volume.

\begin{figure}[!t]
\centering
\includegraphics[width=0.95\linewidth]{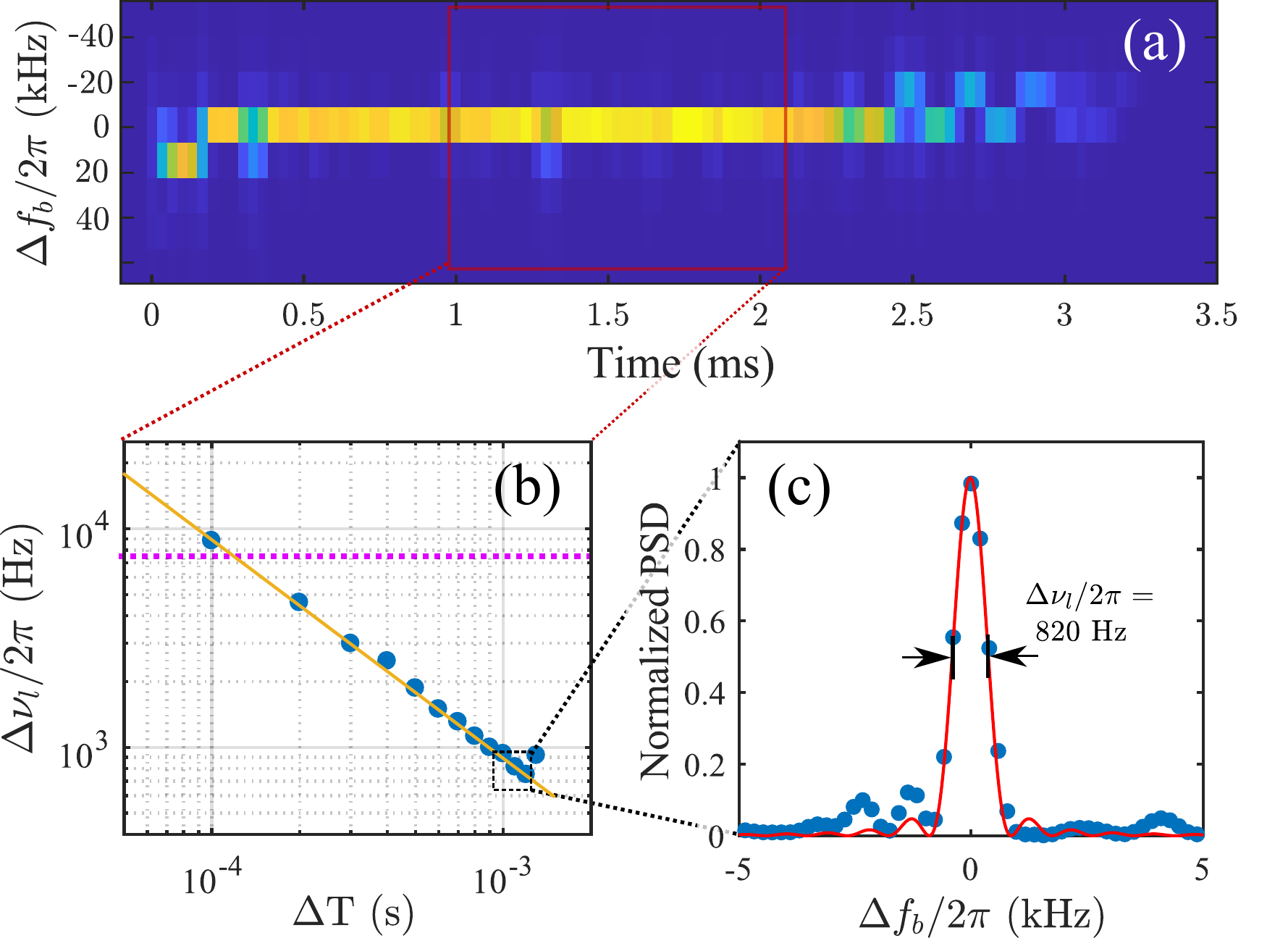} 
\caption{(a) A spectrogram of the measured beat signal $f_b$, with a window size of 60 $\mu$s. The lasing frequency is stable in the middle of the pulse, but drifts towards the end of the pulse when the power drops. 
(b) Finding the FWHM of the PSD of increasingly longer portions of the pulse of the section marked in (a). On top of the data points we plot the Fourier limit of the square signal window. The good agreement suggests that the observed linewidth is Fourier limited. The natural linewidth of 7.5 kHz is indicated with a horizontal dashed purple line. 
(c) The PSD of the beat signal for the steady-state lasing region in (a). The solid red line is the PSD of a Fourier transform of a pure sine burst of constant amplitude with the same center frequency and duration as the beat signal.}
\label{fig4}
\end{figure}

We investigate the frequency contents of the steady-state lasing regime by overlapping the light emitted from the cavity with a frequency stabilized 689 nm reference laser to perform a heterodyne beat measurement. During lasing, the atoms are heated by the repump process, which results in a time-dependent cavity pulling \cite{bohnet2012} of the lasing frequency, $c_p(t)$.
We measure the cavity pulling $c_p(t)$ by producing pulses in the steady-state regime for five different cavity detunings from atomic resonance, $\delta_\text{cav}$. On Fig. \ref{fig3}(a) we show a spectrogram of all five pulses superimposed, where $\Delta f_b$ is the deviation from the mean beat frequency. Each line is the beat signal of a lasing pulse with a different cavity detuning. We determine the peak frequencies belonging to each pulse for each time bin in the spectrogram, fitting a straight line to extract $c_p$ as shown in Fig. \ref{fig3}(b). In Fig. \ref{fig3}(c) we plot $c_p(t)$. We observe that $c_p(t)$ rises rapidly at first and can be approximated by an exponential approach (dashed red). We tentatively assign the increase in cavity pulling during the first 500 $\mu$s of lasing to a rapid increase in the effective sample temperature. The beginning of the lasing pulse is dominated by emission from atoms that absorbed fewer than the average number of repumping photons before reaching the upper lasing level. We make a linear fit to $c_p(t)$ (black line) in the green shaded area to gauge the first order drift in cavity pulling, and find a value of $\Delta c_p(t)= 0.008(3)$ for $\Delta t=1$ ms, which demonstrates low first order sensitivity to the offset of the cavity locking point after the initial 1 ms of lasing.

Next, we investigate the frequency stability of the steady-state lasing regime with the cavity on atomic resonance. Fig. \ref{fig4}(a) shows a spectrogram of the beat signal of a single pulse with overlapping 60 $\mu$s square windows every 30 $\mu$s. We observe lasing regions where the beat frequency, output power and cavity pulling (as seen in Fig. \ref{fig3}(c)) are most stable. We select this region to investigate the linewidth of the beat signal as this period provides steady parameters for the emulation of a continuous steady-state superradiant laser. We estimate the power spectral density (PSD) of the observed beat signal by calculating the absolute square of the discrete Fourier transform, and define the linewidth $\Delta\nu_l$ as the full width at half maximum (FWHM) of the signal peak. 

In Fig. \ref{fig4}(b) we find the PSD for a square signal window of increasingly longer time periods $\Delta T$ within the region shown in Fig. \ref{fig4}(a). We measure the linewidth for each $\Delta T$ by fitting the absolute square of a sinc function, and plot the Fourier limited linewidth of a square signal window $\Delta \nu_l/{2\pi}=0.89/\Delta T$ \cite{riehle} (yellow straight line). If we include more of the pulse in the integration than pictured in Fig. \ref{fig4} (a), $\Delta f_b$ starts to drift due to unstable lasing dynamics and technical noise. The dashed purple line indicates the natural linewidth of the bare lasing transition, $\gamma_l/2\pi$. In Fig. \ref{fig4}(c) we show the PSD of the beat signal (blue circles) of the chosen region of interest in the pulse for a 1.1 ms integration time. The signal has been zero-padded to interpolate the spectrum to a finer resolution, which adds more points to aid the illustration but does not change the measured linewidth. The solid red line is the PSD of a constant amplitude, pure sine burst with the same center frequency and duration as the beat signal. We reach a linewidth of $\Delta\nu_l/{2\pi} = 820$ Hz, almost an order of magnitude lower than the natural linewidth of the transition. The good agreement between the PSD of the beat signal and the pure sine burst suggests that $\Delta\nu_l$ is Fourier limited, which is also evident by how the data points lie on the Fourier limited line in \ref{fig4}(b). While every superradiant pulse we generate exhibits sub-natural linewidth features in the spectrum, the shot-to-shot reproducibility is limited by technical noise of the system.

Our experiment demonstrates sub-natural linewidth light from a single superradiant laser pulse. The continuous repumping scheme combined with carefully adjusted repumping rates allow for pulses long enough to observe steady-state behavior and achieve lasing at technologically relevant powers. Using $^3P_2$ to shelve atoms has interesting prospects for future experiments where the repump process during lasing simultaneously cools the atoms. A MOT centered in an optical cavity operating with $^3P_2$ as the lower state \cite{Grunert2001,Hobson2020}, could get continuously loaded directly from an atomic beam. Then, turning on a weak or detuned 707 nm repump laser would allow for a slow leak of atoms to the upper lasing level to enable continuous superradiant emission. The proof-of-principle in this Letter will inspire further research \cite{thompson2022_1,francesca2022_1,zawada2022_1, jager2021} into fully continuous active atomic clocks, especially with even narrower transitions, such as the $^3P_0$ $\rightarrow$ $^1S_0$ mHz transition in $^{87}$Sr \cite{hotbeam2020,mHzlaser, norcia2018frequency}, which has the potential to compete with state-of-the-art lattice clocks. 

We thank Stefan A. Schäffer, Mikkel Tang and Asbjørn A. Jørgensen for contributions to the experimental apparatus as well as helpful discussions. This project was supported by the European Union’s (EU) Horizon 2020 research and innovation program under the Marie Skłodowska-Curie Grant Agreement No. 860579 (MoSaiQC) and Grant Agreement No. 820404 (iqClock project), the USOQS project (17FUN03) under the EMPIR initiative, and the Q-Clocks project under the European Commission’s QuantERA initiative. We acknowledge funding from VILLUM FONDEN via Research Grant No. 17558. S. L. K. and E. B. contributed equally to this work.
	\bibliographystyle{apsrev4-1}
	\bibliography{biblio_cavity_raman}
	
\end{document}